\documentstyle[aps,preprint]{revtex}
\draft
\begin{document}

\title
{\bf Superconductivity Suppression Close to the Metal-Insulator Transition 
in Strongly Disordered Systems.}

\author{E.Z. Kuchinskii,\ M.V. Sadovskii,\ M.A.Erkabaev}
\address
{Institute for Electrophysics, \\
Russian Academy of Sciences,\ Ural Branch, \\
Ekaterinburg 620049,\ Russia}

\maketitle
\vskip 1.0cm
\begin{center}
{\sl Submitted to JETP,\ November 1996}
\end{center}
\begin{abstract}
On the basis of the recently proposed self-consistent theory of metal-
insulator transition in strongly disordered systems, taking into account
interaction effects, we study transition temperature $T_{c}$ suppression
in disordered superconductors for the wide disorder interval --- from
weakly disordered metal up to Anderson insulator, induced by "Coulomb pseudogap"
formation in the density of states. It is shown that for a number of
systems this theory provides rather satisfactory fit of experimental data.
\end{abstract}


\newpage
\narrowtext

The problem of degradation of superconducting transition temperature under
strong disordering is relatively old\cite{Belitz}. It is closely connected
with the question of superconductivity suppression due to disorder-induced
metal-insulator transition\cite{Sadovskii-I}. A number of microscopic
mechanisms of $T_{c}$ suppression were proposed, such as the growth of Coulomb 
pseudopotential\cite{AMR,LNB}, the influence of Coulomb corrections to the
density of states\cite{Bz} etc. In the majority of papers only small
corrections to $T_{c}$ due to these effects were analyzed. 

Recently we proposed\cite{Kuchinskii-I,Kuchinskii-II} a theory of metal-
insulator transitions which generalize the self-consistent theory of
localization\cite{Vollhardt,Sadovskii-II} taking into account the effects of
electron-electron interaction. This approach has allowed us to study the
behavior of the generalized diffusion coefficient for the wide interval of
disorder parameter both for metallic and insulating regions. 
These results were used in calculations of one-particle density of states
with the account of interelectron interactions. These calculations 
demonstrate the formation and the growth of the "Coulomb pseudogap" in the
density of states close to the Fermi level. In metallic region this behavior
of the density of states corresponds to the usual square-root Altshuler-Aronov
correction\cite{Altshuler}. As disorder parameter grows and system moves
towards the metal-insulator transition this pseudogap deepens, while the
effective region of square-root behavior diminishes, and at the point of the
metal-insulator transition the density of states at the Fermi level becomes
equal to zero --- we obtain a kind of a "Coulomb gap". In the insulating
region, for the band of the finite width, we obtain the typical quadratic
behavior of the density of states close to the Fermi level, reminiscent of
the Coulomb gap due to Efros and Shklovskii\cite{Efros-Shklovskii},\ widening
with the further growth of disorder. Such behavior of the density of states
is in qualitative agreement with experiments on the number of disordered
systems close to the metal-insulator transition\cite{Belitz}, from amorphous
alloys\cite{McMillan,Imry,NbSi-I,NbSi-II} to disordered single-crystals of
metallic oxides, including high-temperature superconductors\cite{Srikanth}.
In this paper the results of these calculations of the density of states
are used for the numerical study of "Coulomb gap" effects on the $T_{c}$
suppression for superconductors which are close to the metal-insulator 
transition.

We shall analyze superconductivity within the framework of the simplest
BCS-model. In the weak coupling approximation the linearized gap-equation
takes the following form\cite{Sadovskii-I}:

\begin{equation}
\Delta(\xi)=-\int\limits_{-\infty}^{\infty} d\xi ' V(\xi,\xi ') N(\xi ')
\frac{1}{2\xi '} th(\frac{\xi '}{2T_{c}}) \Delta(\xi '),
\end{equation}
where $N(\xi)$ - is the averaged on disorder density of states which includes
the effects of electron-electron interaction, $V(\xi,\xi ')$ - is the effective
pairing interaction. The only difference with the standard approach is in the
account of non-trivial dependence of $N(\xi)$ on the electron energy $\xi$, 
close to the Fermi level $E_{F}$.

In BCS theory we usually assume the existence of some effective electron-
electron attraction, which is determined by the balance of pairing attraction
due to electron-phonon interaction and Coulomb repulsion. Thus we consider
the effective pairing interaction in the following simple form:
\begin{equation}
V(\xi,\xi ') = V_{c}(\xi,\xi ') + V_{ph}(\xi,\xi '),
\end{equation} 
where $V_{c}(\xi,\xi ') = V_{c}\theta(E_{F}-|\xi|)\theta(E_{F}-|\xi '|)$ and 
$V_{ph}(\xi,\xi ') = -V_{ph}\theta(\omega_{D}-|\xi|)\theta(\omega_{D}-|\xi '|)$ 
- are the respectively the electron-electron and electron-phonon interactions,
$\omega_{D}$ - is the Debye frequency. The constants $V_{c}>0$ and $V_{ph}>0$
correspond to repulsion and attraction which effectively operate on rather
different intervals of energy:\ $E_{F}\gg \omega_{D}$.

After using this expression in Eq.(1) and some transformations using the
even-odd properties of the gap function $\Delta(\xi)$ we obtain:

\begin{eqnarray}
\Delta(\xi) & = & [ V_{ph}\theta(\omega_{D}-\xi)-V_{c}\theta(E_{F}-\xi) ]
\int\limits_{0}^{\omega_{D}} d\xi ' N(\xi ') \frac{1}{\xi '}
th(\frac{\xi '}{2T_{c}})\Delta(\xi ')- \nonumber \\
& - & V_{c}\theta(E_{F}-\xi) \int\limits_{\omega_{D}}^{E_{F}} d\xi ' 
N(\xi ') \frac{1}{\xi '} th(\frac{\xi '}{2T_{c}})\Delta(\xi ').
\end{eqnarray}

We look for the solution of this equation in the usual two-step form 
\cite{Vonsovskii}:

\begin{equation}
\Delta(\xi) = \left \{ 
\begin{array}{ll}
\Delta_{ph}, & \quad |\xi| < \omega_{D}, \\
\Delta_{c}, & \quad \omega_{D} < |\xi| < E_{F}, \\
\end{array} \right.
\end{equation}
where $\Delta_{ph}, \Delta_{c}$ - are some constants which are determined
from the following system of homogeneous linear equations, which is obtained 
after the substitution of Eq.(4) into Eq.(3):

\begin{eqnarray}
\{1-(V_{ph}-V_{c})N_{0}(0)K(\frac{\omega_{D}}{2T_{c}})\}\Delta_{ph}
+V_{c}N_{0}(0)[K(\frac{E_{F}}{2T_{c}})-K(\frac{\omega_{D}}{2T_{c}})]\Delta_{c} 
= 0, \nonumber \\
V_{c}N_{0}(0)K(\frac{\omega_{D}}{2T_{c}})\Delta_{ph}+\{1+V_{c}N_{0}(0)
[K(\frac{E_{F}}{2T_{c}})-K(\frac{\omega_{D}}{2T_{c}})]\Delta_{c} = 0,
\end{eqnarray}
where $N_{0}(0)$ - is the one-electron density of states of noninteracting
electrons at the Fermi level and we introduce

\begin{equation}
K(\xi)=\int\limits_{0}^{\xi}d\xi '\frac{1}{\xi '}th(\xi ')
\left [{N(2T_{c}\xi ')} \over {N_{0}(0)} \right].
\end{equation}
Equation for $T_{c}$ follows from the usual zero-determinant condition for
this homogeneous system:

\begin{eqnarray}
(\lambda & - & \mu^{\ast})K(\frac{\omega_{D}}{2T_{c}})=1, \nonumber \\
\mu^{\ast} = \mu\{
1 & + & \mu[K(\frac{E_{F}}{2T_{c}})-K(\frac{\omega_{D}}{2T_{c}})]\}^{-1},
\end{eqnarray}
where $\mu^{\ast}$ - is the Coulomb pseudopotential, $\mu=V_{c}N_{0}(0)$ 
- is the Coulomb constant, $\lambda=V_{ph}N_{0}(0)$ - is the pairing constant
due to electron-phonon interaction. In the clean limit, when the density of
states at the Fermi level is constant, this reduces to the usual BCS-expression
for $T_{c}$. 

Equation (7) for $T_{c}$ was solved numerically for disorder parameter
changing in wide interval both for metallic and insulating regions. The
density of states was calculated taking into account lowest order corrections
over electron-electron interaction\cite{Kuchinskii-I,Kuchinskii-II}:

\begin{equation}
N(\xi)=-\frac{1}{\pi}Im\int\frac{d^{3}{\bf p}}{(2 \pi)^{3}}
G^{R}({\bf p},\xi),
\end{equation}
where $G^{R(A)}({\bf p},\xi)=
[\xi-\xi_{p}\pm i\gamma-\Sigma^{R(A)}_{ee}({\bf p},\xi)]^{-1}$ 
- is the retarded (advanced) electron Green's function, 
$\Sigma^{R(A)}_{ee}({\bf p},\xi)$ - is "Fock" contribution to electron self-
energy\cite{Kuchinskii-I,Altshuler}:

\begin{equation}
\Sigma^{R(A)}_{ee}({\bf p},\xi) \approx 
4i\gamma^{2}\mu N^{-1}_{0}(0)
G^{A(R)}_{0}({\bf p},\xi)
\int\limits_{\xi}^{\infty}\frac{d\omega}{2\pi}
\int\limits_{|{\bf q}|<k_{0}}\frac{d^{3}{\bf q}}{(2\pi)^{3}}
\frac{1}{[-i\omega+D(\omega)q^{2}]^{2}}.
\end{equation}
Here $D(\omega)$ - is the generalized diffusion coefficient, which is determined
from the following self-consistent nonlinear integral equation
\cite{Kuchinskii-I,Kuchinskii-II}:
\begin{eqnarray}
\frac{D(\omega)}{D_{0}} & = & 1-\frac{1}{\pi N_{0}(0)}\frac{D(\omega)}{D_0}
\int\limits_{|{\bf q}|<k_{0}} \frac{d^{3}{\bf q}}{(2 \pi)^{3}}
\frac{1}{-i\omega+D(\omega)q^{2}}+ \nonumber\\
& + & \frac{8i}{3\pi}\frac{\mu D_{0}}{\pi N_{0}(0)}
\int\limits_{\omega}^{\infty} d\Omega
\int\limits_{|{\bf q}|<k_{0}} \frac{d^{3}{\bf q}}{(2 \pi)^{3}}
\frac{q^{2}}{(-i(\Omega+\omega)+D(\Omega+\omega)q^{2})
(-i\Omega+D(\Omega)q^{2})^{2}},
\end{eqnarray}
where $D_{0}=E_{F}/\ 3m\gamma$ - is the usual Drude diffusion coefficient, 
$\gamma=1/2\tau$ - Born scattering rate, $\tau$ - mean free time, 
$k_{0}= min\{p_{F},l^{-1}\}$ - cut-off in the momentum space, $p_{F}$ - 
Fermi momentum, $l$ - mean free path. The data on static conductivity used below
were also obtained by numerical solution of Eq.(10)
\cite{Kuchinskii-I,Kuchinskii-II}.

In Fig.1 we show the behavior of the density of states close to the Fermi
level which demonstrates the evolution of the "Coulomb pseudogap" as
disorder grows. This behavior obviously leads to superconducting $T_{c}$
suppression as the system moves towards the metal-insulator transition.

Fig.2 demonstrates $T_{c}$ suppression as disorder parameter $(p_{F}l)^{-1}$ 
grows for different values of the Coulomb constant $\mu$ and fixed value of
pairing constant $\lambda$. For large $\mu$ and growing disorder 
$(p_{F}l)^{-1}$ the value of $T_{c}$ drops rather fast and becomes zero in
metallic region far enough from metal-insulator transition. 
For smaller values of $\mu$ this drop of $T_{c}$ with growing disorder 
$(p_{F}l)^{-1}$ becomes slower and for small $\mu$ and large enough 
$\lambda$ (dashed curves) we get the possibility of superconductivity
persisting even in the insulating region\cite{Sadovskii-I}. This possibility
is clearly demonstrated at the insert in Fig.2, where we show the $T_{c}$
dependence on the static conductivity of the system $\sigma$ for the
appropriate values of $\lambda$ and $\mu$. For large values of $\mu$ 
as conductivity $\sigma$ drops $T_{c}$ also drops and superconductivity is
completely suppressed rather far from the metal-insulator transition. 
For small values of $\mu$ this drop of $T_{c}$ with $\sigma$ becomes slower
and for sufficiently large values of $\lambda$ (dashed curves) $T_{c}$
remains finite even in the case of $\sigma \rightarrow 0$.

Fig.3 demonstrates $T_{c}$ degradation with the growth of disorder parameter
$(p_{F}l)^{-1}$  for different values of the pairing constant $\lambda$ for
the fixed value of the Coulomb constant $\mu$. For small $\lambda$ and
disorder parameter $(p_{F}l)^{-1}$ growing the value of $T_{c}$ drops rather
fast and becomes zero in the metallic state far from the metal-insulator 
transition. As $\lambda$ grows this drop of $T_{c}$ becomes slower and for
large enough values of $\lambda$ superconductivity is completely suppressed
only somewhere in the insulating region. At the insert in Fig.3 we show the
dependence of the Coulomb pseudopotential $\mu^{\ast}$ on the disorder
parameter $(p_{F}l)^{-1}$ for the appropriate values of $\lambda$ and $\mu$
demonstrating rather insignificant growth of $\mu^{\ast}$ with disorder 
$(p_{F}l)^{-1}$ close to the point where superconductivity is completely
suppressed. Apparently this behavior is natural enough because we neglect
all the processes renormalizing the matrix element of Coulomb interaction in
Eq.(2) due Anderson localization and electron-electron interactions which
can actually lead to rather important growth of Coulomb pseudopotential
close to the metal-insulator transition \cite{Sadovskii-I}. 

This kind of behavior of $T_{c}$ on the static conductivity $\sigma$ and on
disorder parameter was observed experimentally in a number of disordered
systems, which remain superconducting close to the disorder induced metal-
insulator transition\cite{Belitz}, \cite{Sadovskii-I}, 
\cite{McMillan} - \cite{Srikanth}, \cite{AuSi-I} - \cite{AuSi-III}.
Our results agree rather well with experiments on amorphous alloys of
$InO_{x}$ \cite{InOx}, $Nb_{x}Si_{1-x}$ 
\cite{NbSi-I,NbSi-II}, $Au_{x}Si_{1-x}$ \cite{AuSi-I,AuSi-II,AuSi-III}. 

The authors of Ref.\cite{InOx} had presented the results of the measurements of
disorder parameter $(p_{F}l)^{-1}$ for the amorphous alloy of $InO_{x}$,\ 
as well as the data for $T_{c}$ and static conductivity close to the metal-
insulator transition. According to our work\cite{Kuchinskii-I,Kuchinskii-II} 
the static conductivity close to the metal-insulator transition can be
expressed as:

\begin{equation}
\sigma=\sigma_{0}[(p_{F}l)W_{c}(\mu)-1],
\end{equation}
where $\sigma_{0}$ - is some characteristic scale of conductivity close to the
metal-insulator transition, $W_{c}(\mu)$ - the value of disorder parameter
(depending on the Coulomb constant) corresponding to the point of metal-
insulator transition. Approximating the experimental data for
$InO_{x}$ by Eq.(11) allows us to estimate the characteristic conductivity
scale $\sigma_{0}$ and also, from the value of $W_{c}$, the Coulomb constant
$\mu$. Satisfactory correlation (Cf. the insert in Fig.3) are obtained for
the following values:
$\sigma_{0} \simeq 324.95$ $\Omega^{-1} \cdot cm^{-1}$, 
and $W_{c} \simeq 0.606$ giving $\mu \simeq 1.0$. 

Fig.3 demonstrates the comparison of our results with experimental data
on $T_{c}$ dependence on static conductivity $\sigma$ for the amorphous
$InO_{x}$ using the value of $T_{co} = 3.41$ $K$, 
$\omega_{D} = 112$ $K$ and $E_{F} = 9.98 \cdot 10^{4}$ $K$, 
$[\omega_{D}/E_{F}] \simeq 1.1\cdot10^{-3}$ - for pure $In$ and the given
above values of $\sigma_{0}$ and $\mu$, which allows to estimate the
pairing constant $\lambda$. Satisfactory agreement is obtained for
$\lambda \simeq 0.45$. Dashed curves correspond to the values of 
$\lambda \simeq 0.4$ and $0.5$.

Let us discuss now the results for $T_{c}$ and static conductivity 
dependence on the $Si$ content for amorphous alloys of $Nb_{x}Si_{1-x}$ 
\cite{NbSi-I,NbSi-II} and $Au_{x}Si_{1-x}$ \cite{AuSi-I,AuSi-II,AuSi-III} 
close to the metal-insulator transition. Assuming that the disorder parameter
in this case is just proportional to $Si$ concentration,\ so that 
$(p_{F}l)^{-1} \sim 1-x$, we can express Eq.(11) for static conductivity in
the following form: 

\begin{equation}
\sigma=\sigma_{0}\frac{x-x_{c}}{1-x},
\end{equation}
where $x_{c}$ - the critical concentration of $Nb$ or $Au$ at the point of
metal-insulator transition. Approximating the experimental data for
conductivity in $Nb_{x}Si_{1-x}$ ¨ $Au_{x}Si_{1-x}$ by Eq.(12) 
allows us to estimate $\sigma_{0}$ and critical concentration $x_{c}$. 
Satisfactory correlation (Cf. inserts in Fig.5 and Fig.6) is obtained for:
\\
$Nb_{x}Si_{1-x}$: $\sigma_{0} \simeq 1963.9$ $\Omega^{-1} \cdot cm^{-1}$, 
$x_{c} \simeq 0.115$; \\
$Au_{x}Si_{1-x}$: 
$\sigma_{0} \simeq 2782.13$ $\Omega^{-1} \cdot cm^{-1}$, 
$x_{c} \simeq 0.14$.

Fig.5 and Fig.6 present the comparison our results with the experimental data
for $T_{c}$ - dependence on conductivity $\sigma$ for amorphous 
$Nb_{x}Si_{1-x}$ and $Au_{x}Si_{1-x}$, using for the pure $Nb$:
$T_{co} = 9.26$ $K$, $\omega_{D} = 276$ $K$ and $E_{F} = 6.18 \cdot 10^{4}$ $K$, 
$[\omega_{D}/E_{F}] \simeq 3.0\cdot10^{-3}$; 
while for $Au_{x}Si_{1-x}$ we assume $T_{co} = {T_{c}}_{max} \simeq 0.86$ $K$, 
$\omega_{D} = 170$ $K$ and $E_{F} = 6.42 \cdot 10^{4}$ $K$, 
$[\omega_{D}/E_{F}] \simeq 0.9\cdot10^{-3}$ 
with the mentioned above values of $\sigma_{0}$, which allows us to estimate the
pairing constant $\lambda$. Assuming for these systems the Coulomb constant
$\mu \simeq 1$, the satisfactory agreement is obtained in case of
$Nb_{x}Si_{1-x}$ for $\lambda \simeq 0.54$ and for $Au_{x}Si_{1-x}$ with 
$\lambda \simeq 0.62$.

Surely the results presented above are essentially based upon the simplest
BCS-model and are probably oversimplified. More rigorous approach to calculations
of $T_{c}$ must be based upon Eliashberg equations and realistic models of
electron-phonon interaction\cite{Vonsovskii}. Especially this is important in
case of large enough values of $\lambda$, which demonstrate the possibility of
superconductivity persisting in the insulating phase. At the same time, in
present paper we were not concerned with the problem of the genesis of the
initial value of $T_{c0}$ in pure system, but were studying only the $T_{c}$
dependence on disorder. In this sense our results may be also qualitatively
valid also in the case of strong-coupling superconductivity. We must also note
that the more rigorous analysis is also needed taking into account disorder
effects in the matrix element of Coulomb repulsion, which lead to the
additional mechanism of $T_{c}$ degradation $T_{c}$ \cite{Sadovskii-I,AMR,LNB}.
In the present work we have only taken into account pseudogap effects in
the density of states. It is possible that rather satisfactory agreement with
experiments can signify the dominating role of pseudogap formation effects
in the problem of $T_{c}$ degradation under disordering, which was claimed
(on the level of small corrections) already in Ref.\cite{Bz}.

This work was performed with the partial support of Russian Foundation of
Basic Research under the grant No.96-02-16065,\ as well as under the Project
IX.I of the Program "Statistical Physics" of the Russian Ministry of Science.

\newpage

\begin{center}
{\bf\large Figure Captions:}
\end{center}

{\bf Fig.1.} Density of states in the case of band of finite width
$2E_{F}$ for $\frac{8}{3\pi}\mu=1.0$ for the different values of disorder
parameter $(p_{F}l)^{-1}$: 1 - 0.1,..., 5 - 0.5 - in metallic region, 
7 - 0.7,..., 10 - 1.0 - in insulating region. Dashed curve - 6
corresponds to the point of the metal-insulator transition. Energy 
$\varepsilon$ is in the units of $D_{0}k_{0}^{2}$.

{\bf Fig.2.} $T_{c}$ degradation as a function of disorder parameter
$(p_{F}l)^{-1}$ for the fixed value of the pairing constant $\lambda$ 
($\lambda=0.5$ - full curves, $\lambda=1.0$ - dashed curves) for the 
different values of Coulomb constant $\frac{8}{3\pi}\mu$: 
1 - 0.2, ..., 5 - 1.0. At the insert:\ $T_{c}$ dependence upon the static
conductivity $\sigma$ for the appropriate values of pairing constant 
$\lambda$ and Coulomb constant $\mu$. 

{\bf Fig.3.} $T_{c}$ degradation as a function of disorder parameter
$(p_{F}l)^{-1}$ for the fixed value of Coulomb constant
$\frac{8}{3\pi}\mu=0.4$ and different values of the pairing constant
$\lambda$: 1 - 0.3, 2 - 0.4, ..., 8 - 1.0. At the insert:\ dependence of
Coulomb pseudopotential $\mu^{\ast}$ on disorder parameter
$(p_{F}l)^{-1}$ for the appropriate values of pairing constant $\lambda$ 
and Coulomb constant $\mu$. The arrow shows the point of the metal-insulator
transition.

{\bf Fig.4.} $T_{c}$ dependence on conductivity $\sigma$ for the amorphous
alloys of $InO_{x}$. At the insert:\ approximation of conductivity
dependence upon the disorder parameter $(p_{F}l)^{-1}$.

{\bf Fig.5.} $T_{c}$ dependence on conductivity $\sigma$ for the amorphous
alloys of $Nb_{x}Si_{1-x}$. At the insert:\ approximation of conductivity
dependence upon concentration of $Nb$.

{\bf Fig.6.} $T_{c}$ dependence on conductivity $\sigma$ for the amorphous
alloys of $Au_{x}Si_{1-x}$. At the insert:\ approximation of conductivity
dependence upon concentration of $Au$.

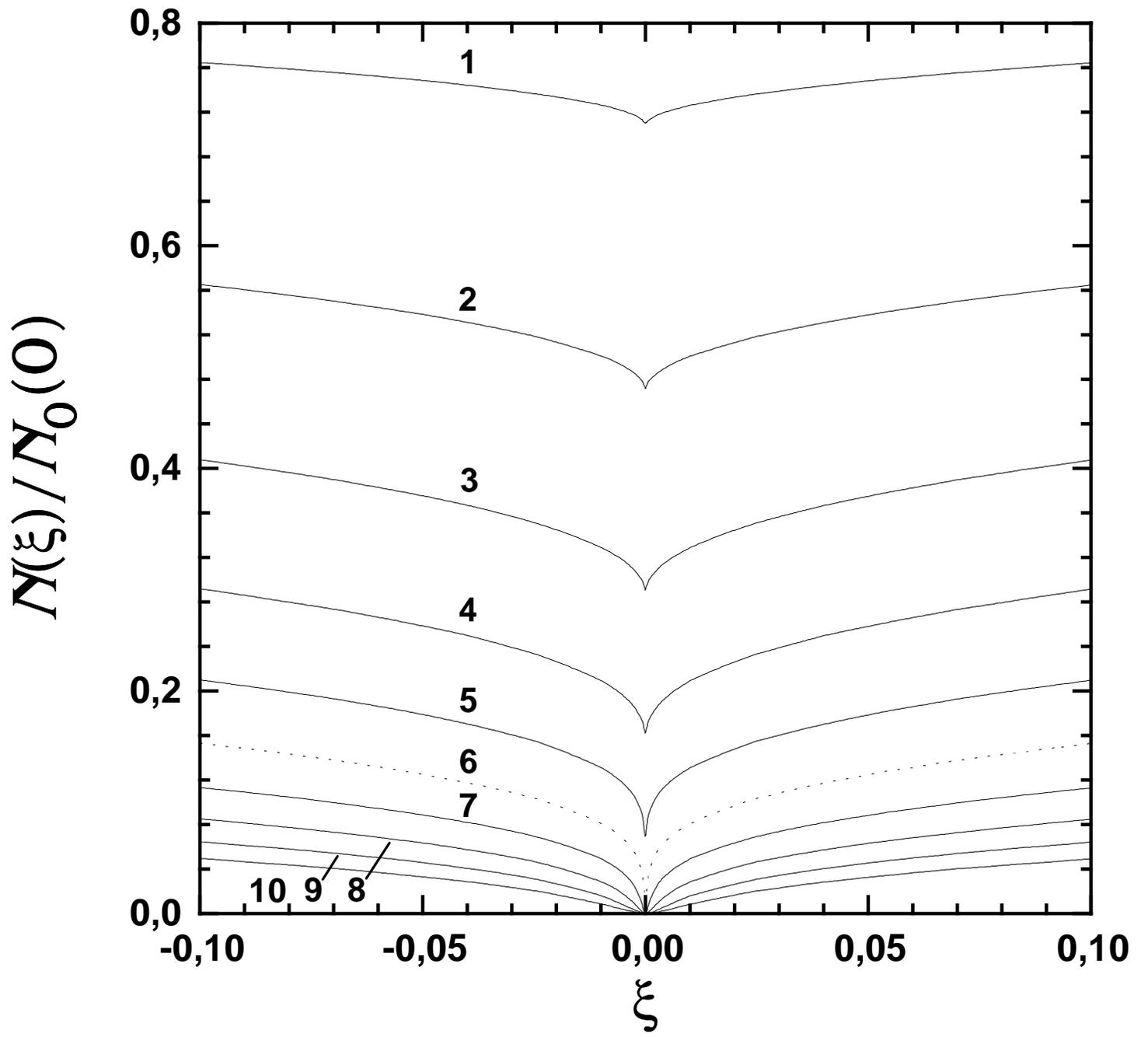
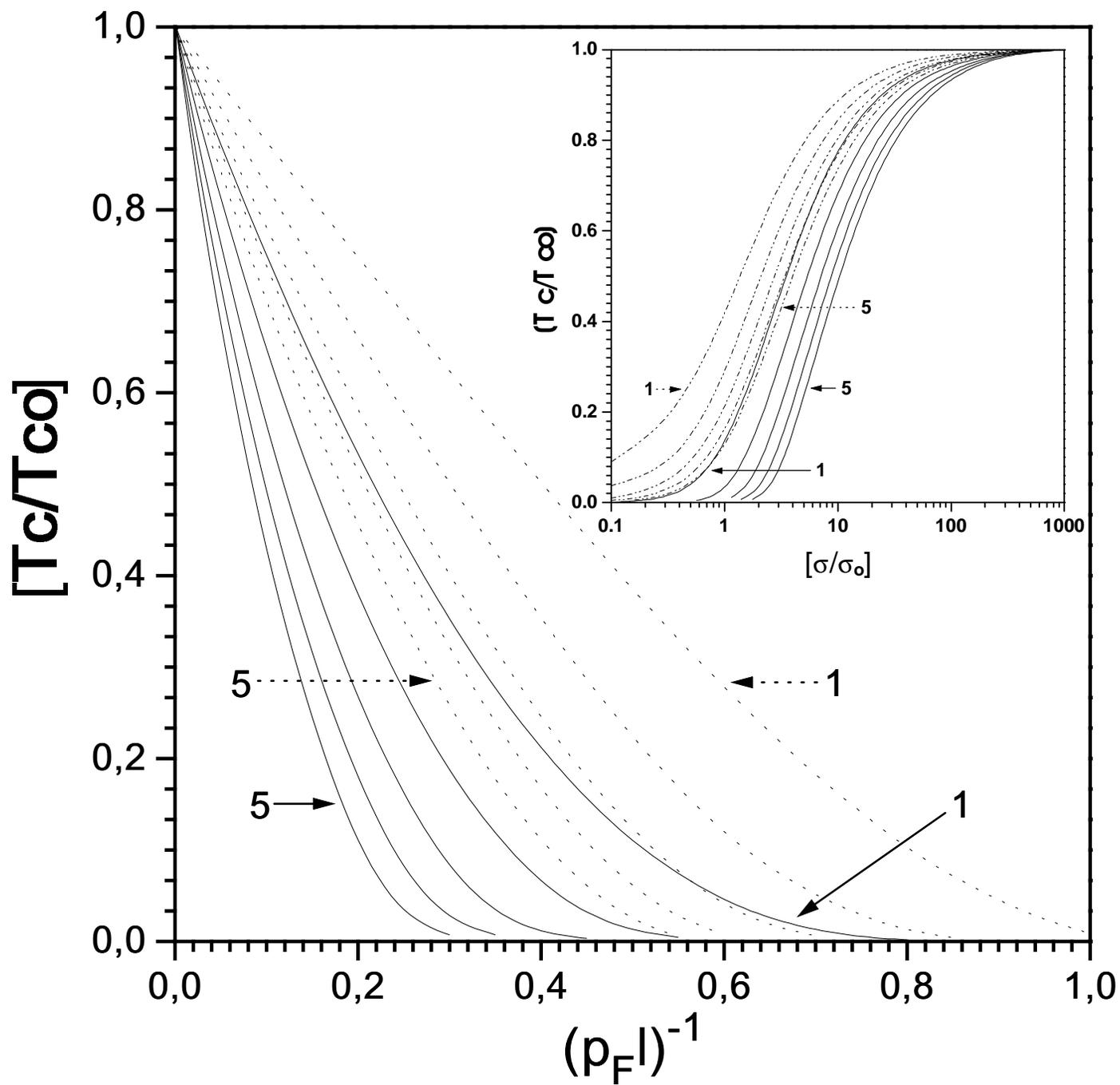
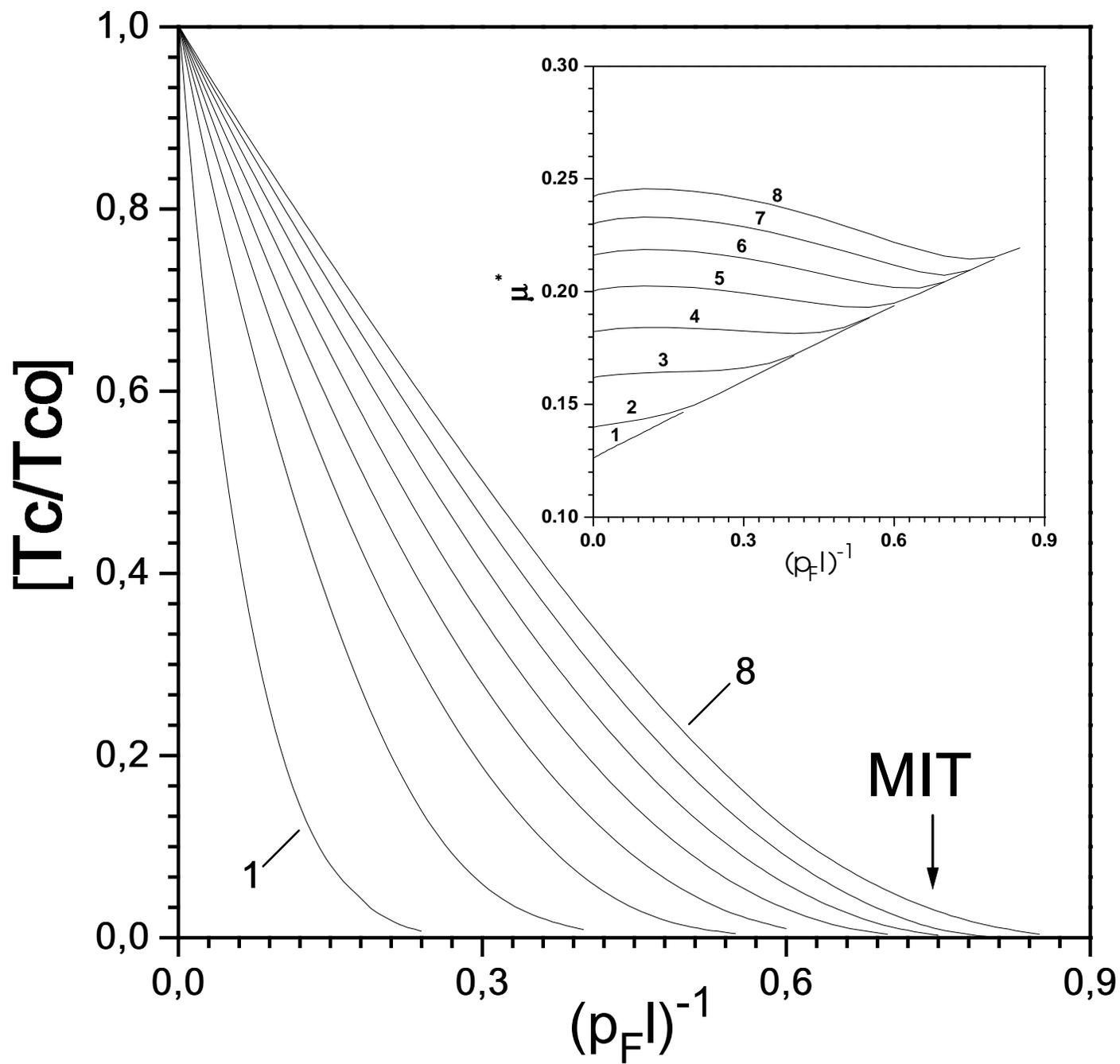
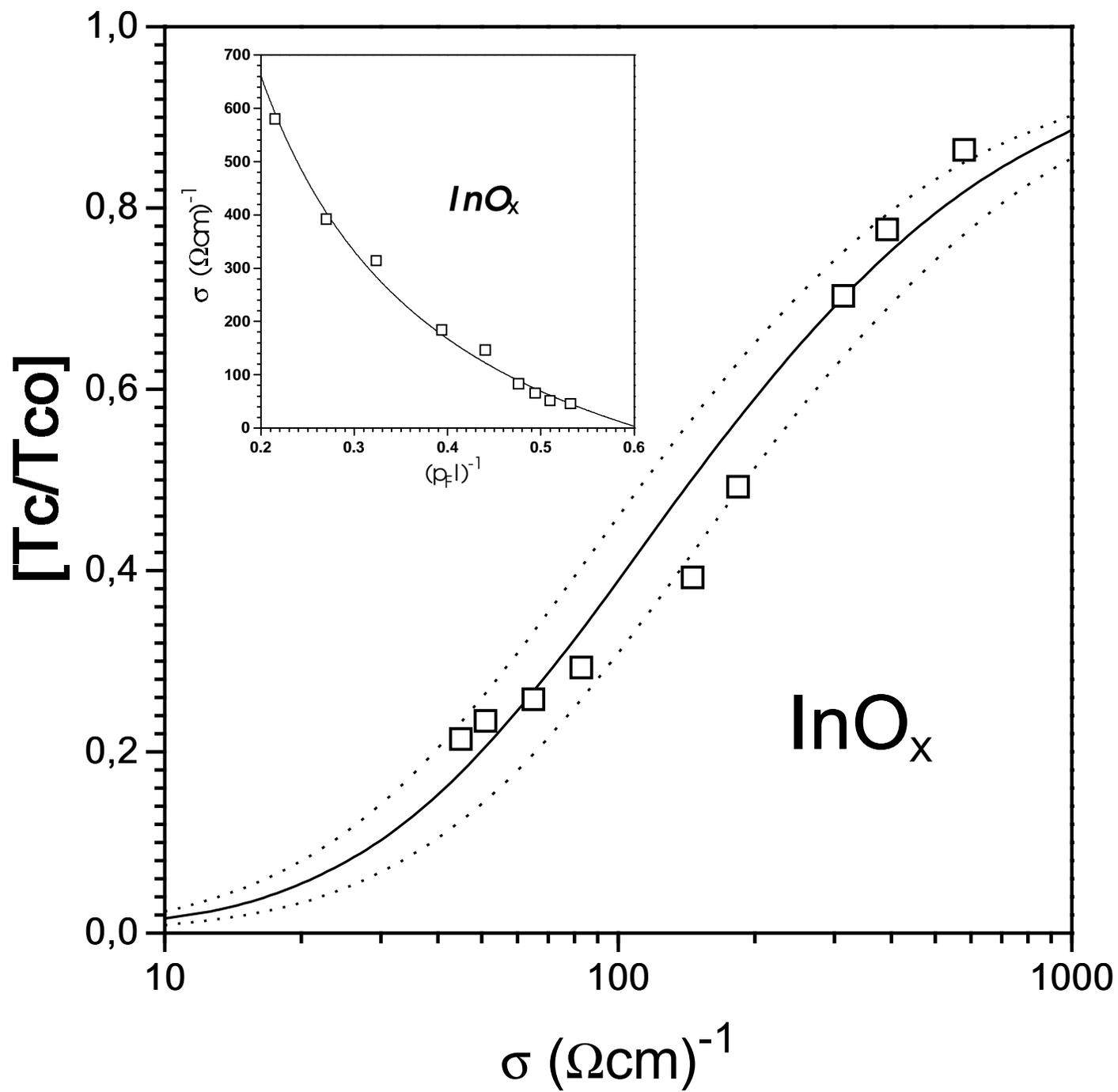
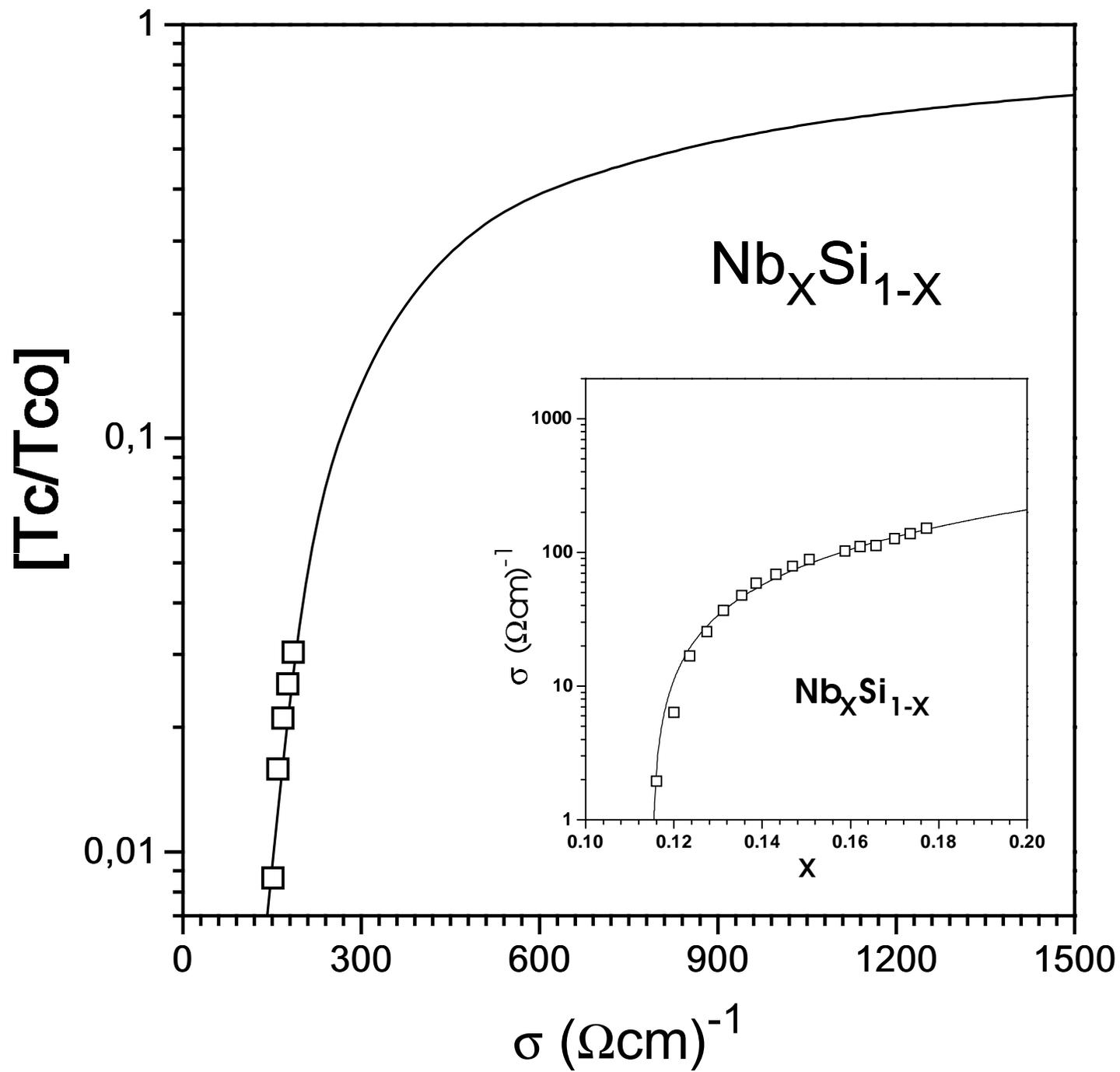
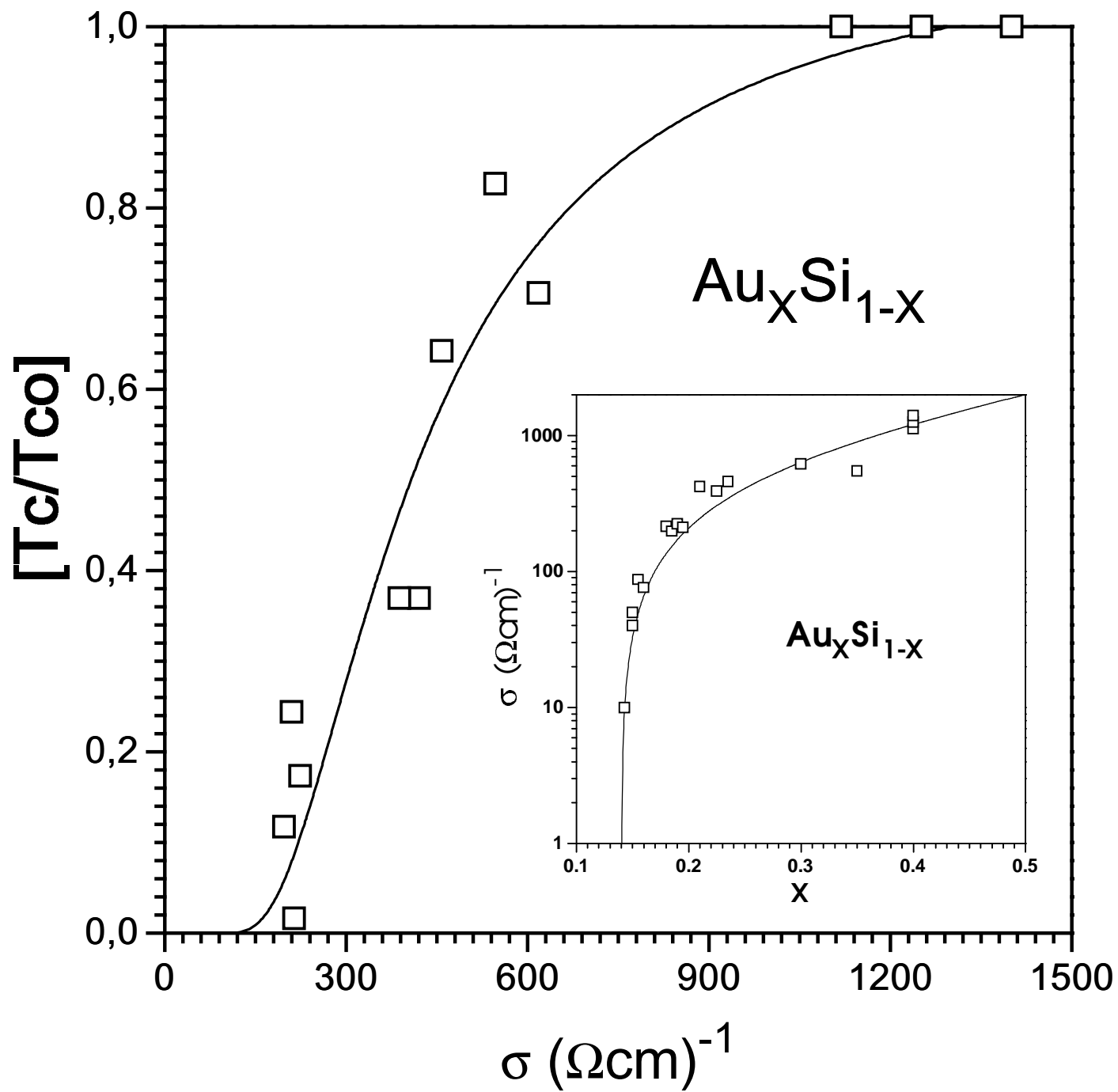
\end{document}